\documentstyle[sprocl]{article} 

\def\Journal#1#2#3#4{{#1} {\bf #2}, #3 (#4)} 
 
\def\NPB{{\em Nucl. Phys.} B} 
\def\PLB{{\em Phys. Lett.}  B} 
\def\PRL{\em Phys. Rev. Lett.} 
\def\PRD{{\em Phys. Rev.} D} 
 
\def\be{\begin{equation}} 
\def\ee{\end{equation}} 
\def\bea{\begin{eqnarray}}
\def\eea{\end{eqnarray}}
\def\ba{\begin{array}}
\def\ea{\end{array}}
 
\def\simge{\mathrel{%
   \rlap{\raise 0.511ex \hbox{$>$}}{\lower 0.511ex \hbox{$\sim$}}}}
\def\simle{\mathrel{
   \rlap{\raise 0.511ex \hbox{$<$}}{\lower 0.511ex \hbox{$\sim$}}}}
 
\def\slashchar#1{\setbox0=\hbox{$#1$}           
   \dimen0=\wd0                                 
   \setbox1=\hbox{/} \dimen1=\wd1               
   \ifdim\dimen0>\dimen1                        
      \rlap{\hbox to \dimen0{\hfil/\hfil}}      
      #1                                        
   \else                                        
      \rlap{\hbox to \dimen1{\hfil$#1$\hfil}}   
      /                                         
   \fi}                                         %
\def\ts{\thinspace}

\def\ra{\rightarrow}

\def\Lra{\Longrightarrow}

\def\llra{\longleftrightarrow}
\def\ol{\bar}

\def\CH{{\cal H}}

\def\CO{{\cal O}}

\def\suc{SU(3)}

\def\Ntc{N}
\def\sutc{SU(N)}
\def\Getc{G_{ETC}}

\def\uone{U(1)_1}
\def\utwo{U(1)_2}
\def\uy{U(1)_Y}
\def\suone{SU(3)_1}
\def\sutwo{SU(3)_2}

\def\kslash{\raise.15ex\hbox{/}\kern-.57em k}

\def\condtbt{\langle \bar t t\rangle}

\def\condaa{\langle \bar T^1_L T^1_R\rangle}
\def\condbb{\langle \bar T^2_L T^2_R\rangle}
\def\condab{\langle \bar T^1_L T^2_R\rangle}

\def\condij{\langle \bar T^i_L T^j_R\rangle}

\def\toppip{\pi_t^+}

\def\gev{{\rm GeV}}
\def\tev{{\rm TeV}}

\def\half{{\textstyle{ { 1\over { 2 } }}}}
\def\third{{\textstyle{ { 1\over { 3 } }}}}
\def\fourth{{\textstyle{ { 1\over { 4 } }}}}

\def\twothirds{{\textstyle{ { 2\over { 3 } }}}}
\def\sixth{{\textstyle{ { 1\over { 6 } }}}}

\begin{document} 

\hskip3.4truein\hbox{BUHEP-97-8}

\hskip3.4truein\hbox{hep-ph/9703233}

\vskip0.3truein

\title{PROGRESS ON SYMMETRY BREAKING AND GENERATIONAL MIXING IN
TOPCOLOR-ASSISTED TECHNICOLOR}
 
\author{ KENNETH LANE } 
 
\address{\hskip0.6truein Department of Physics, Boston
  University,\hfil\break
590 Commonwealth Ave, Boston, MA 02215, USA}

\maketitle\abstracts{Topcolor-assisted technicolor provides a dynamical
explanation for electroweak and flavor symmetry breaking and for the large
mass of the top quark without unnatural fine tuning.  I briefly review the
basics of topcolor-assisted technicolor, including major constraints and a
general approach to satisfying them. The main challenge to
topcolor-assisted technicolor is to generate the observed mixing between
heavy and light generations while breaking the strong topcolor interactions
near~$1\,\tev$. I argue that these phenomena, as well as electroweak
symmetry breaking, are intimately connected and I present a scenario for
them based on nontrivial patterns of technifermion condensation. I also
exhibit a class of models realizing this scenario.}

\section{Introduction}\label{sec:intro}
 
This is the written version of my talk at the 1996 Workshop in Nagoya on
Strongly Coupled Gauge Theories (SCGT~96) on the status of models of
topcolor-assisted technicolor (TC2).~\footnote{A few of the remarks here
will be new, reflecting work done since the SCGT~96 Workshop.} This topic
is in its infancy, and I believe that much development lies ahead. Thus,
this report must be viewed as preliminary. At the same time, I hope it will
stimulate study and resolution of many of the problems that TC2 faces.

This report has the following plan: In Sec.~\ref{sec:hill}, I review the
basics of the ``simplest'' TC2 model discussed by Chris Hill at this
workshop. Then, in Sec.~\ref{sec:constraints}, I summarize the principal
constraints on this early version of TC2 model-building. The response to
these constraints is outlined in Sec.~\ref{sec:natural}. This response did
not resolve satisfactorily the mechanism by which the third generation
quarks mix with those of the first two generations. It touched not at all
on the question of how the color and hypercharge groups of the heavy and
light generations break down to their diagonal subgroup, the familiar
$SU(3) \otimes \uy$. These problems are addressed in Sec.~\ref{sec:heart};
the discussion there is the heart of this paper. As I said, this work
is still very much under development.

\section{Review of Hill's Simplest Topcolor-Assisted Technicolor
Model}\label{sec:hill}

Topcolor-assisted technicolor (TC2) was proposed by
Hill~\cite{tctwohill}$^{,\ts}$\cite{hillscgt} to overcome shortcomings of
top-condensate models of electroweak symmetry
breaking~\cite{topcondref}$^{,\ts}$\cite{topcref} and of technicolor models
of dynamical electroweak and flavor symmetry
breaking.~\cite{tcref}$^{,\ts}$\cite{etcsdls}$^{,\ts}$\cite{etceekl}
Technicolor and extended technicolor (ETC) have been unable to provide a
natural and plausible understanding of why the top quark mass is so
large.~\cite{toprefs} On the other hand, models in which strong topcolor
interactions drive top-quark condensation {\it and} electroweak symmetry
breaking are unnatural. To reproduce the one-Higgs-doublet standard model
consistent with precision electroweak measurements (especially of the
parameter $\rho = M_W^2/M_Z^2 \cos^2\theta_W \simeq 1$), the topcolor
energy scale must be much greater than the electroweak scale of
$\CO(1\,\tev)$. This requires severe fine tuning of the topcolor coupling.

Hill's combination of topcolor and technicolor keeps the best of both
schemes. In TC2, technicolor interactions at the scale $\Lambda_{TC} \simeq
\Lambda_{EW} \simeq 1\,\tev$ are mainly responsible for electroweak
symmetry breaking. Extended technicolor is still required for the hard
masses of all quarks and leptons {\it except} the top quark. Topcolor
produces a large top condensate, $\condtbt$, and all but a few GeV of $m_t
\simeq 175\,\gev$. This remaining portion of $m_t$ {\it must} be generated
by ETC interactions in order that the Nambu-Goldstone
bosons---top-pions---associated with top condensation acquire appreciable
masses. Hill has pointed out that some, perhaps all, of the bottom quark
mass may arise from $\suone$ instantons.~\cite{tctwohill} Top condensation
contributes comparatively little to electroweak symmetry breaking. Thus,
the scale of topcolor can be lowered to near~$1\,\tev$ and the interaction
requires little or no fine tuning.

In Hill's simplest TC2 model,\cite{tctwohill} there are separate color and
weak hypercharge gauge groups for the heavy third generation of quarks and
leptons and for the two light generations. He assumed that the third
generation transforms under {\it strongly-coupled} $\suone \otimes \uone$
with the usual charges, while the light generations transform in the usual
way under {\it weakly-coupled} $\sutwo \otimes \utwo$. By some as yet
unspecified mechanism, these four groups are broken near the electroweak
energy scale of about $1\,\tev$ to the diagonal subgroup of ordinary color
and hypercharge, $\suc\otimes \uy$. The desired pattern of condensation for
the third generation occurs because the $\uone$ couplings are such that the
spontaneously broken $\suone \otimes \uone$ interactions are supercritical
for the top quark, but not for the bottom quark (and certainly not for the
tau lepton).

Nothing is said in this scenario about how topcolor breaks. Nor is there
any mention of how third generation quarks mix with and, hence, decay to
those of the first two generations. Topcolor-assisted technicolor is
natural in the scenario presented so far. The question is: can its
naturalness be maintained in a more realistic model, one that accounts for
topcolor breaking and generational mixing?

\section{Constraints on Topcolor-Assisted
Technicolor}\label{sec:constraints}

Two important constraints were imposed on TC2 soon after Hill's proposal
was made. The first is due to Chivukula, Dobrescu and Terning
(CDT).~\cite{cdt} A small part of the top-quark mass must arise from ETC
interactions and CDT assumed, plausibly, that at least some of the
bottom-quark mass does as well. If, as CDT further assumed, these hard
masses arise from $(t,b)$ couplings to the {\it same} doublet of
technifermions, then the latter must have custodial-isospin violating
couplings to the strong $\uone$. To keep $\rho \simeq 1$, they then showed
that the $\uone$ interaction must be so weak that it is necessary to
fine-tune the $\suone$ coupling to within~1\% of its critical value for top
condensation {\it and} to increase the topcolor boson mass above
$4.5\,\tev$. Thus, TC2 seemed to be unnatural after all. CDT did state
that their bounds could be relaxed if $\uone$ couplings did not violate
isospin. However, they expected that this would be difficult to
implement because of the requirements of canceling gauge anomalies and
of allowing mixing between the third and first two generations. As we
shall see, the difficult issue will be the intergenerational mixing.

The second constraint on TC2 is due to Kominis.~\cite{kominis} He pointed
out that, under the likely assumption that the $b$-quark's topcolor
interactions are not far from critical, there will be relatively light,
$M \simeq 250$-$350\,\gev$, scalar bound states of $\ol t_L b_R$ and $\ol
b_L b_R$. These scalars couple strongly ($\propto m_t$) to third generation
quarks. Thus, they can induce potentially large $B_d-\ol B_d$ mixing.
Kominis showed that the measured value of $\Delta M_{B^0_d}/M_{B^0_d}$
implies the upper bound
\be\label{dmixing}
|D^d_{Lbd} D^d_{Lbb} D^d_{Rbd} D^d_{Rbb}| \simle 10^{-7}
\ee
on elements of the unitary matrices which diagonalize the (generally
nonhermitian) $Q=-\third$ quark mass matrix. If, as we shall make plausible
later, the elements of the Kobayashi-Maskawa (KM) matrix connecting the
third generation to the first two arise mainly from mixing in the $Q =
-\third$ sector, then this limit can be compared to $|D_{Lbd}| \simeq
|V_{ub}| \simeq 0.002$--0.005, while $|D_{Lbb}| \simeq 1$. Mixing will have
to be extraordinarily small in the right-handed-down sector if Kominis'
constraint is to be satisfied.

\begin{table}[t]
\caption{Quark and technifermion hypercharges and electric
charges in the TC2 models of Ref.~[11]. The $\uone$ hypercharges $x_i, y_i,
z_i$ are given in the text.
\label{tab:tabl1}}
\vspace{0.4cm}
\begin{center}
\begin{tabular}{|c|c|c|c|}
\hline
Particle & 
$Y_1$ & 
$Y_2$ & 
$Q = T_3 + Y_1 + Y_2$ \\
\hline
\hline
$q_L^l$ & $0$ & $\sixth$ & $\twothirds$, $-\third$ \\
$c_R$, $u_R$ & $0$ & $\twothirds$ & $\twothirds$ \\
$d_R$, $s_R$ & $0$ & $-\third$ & $-\third$ \\
\hline
$q_L^h$ & $\sixth$ & $0$ & $\twothirds$, $-\third$ \\
$t_R$ & $\twothirds$ & $0$& $\twothirds$ \\
$b_R$ & $-\third$  & $0$ & $-\third$ \\
\hline
$T_L^l$ & $x_1$ & $x_2$ & $\pm\half + x_1 + x_2$ \\
$U_R^l$ & $x_1$ & $x_2+\half$ & $\half + x_1 + x_2$ \\
$D_R^l$ & $x_1$ & $x_2-\half$ & $-\half + x_1 + x_2$ \\
\hline
$T_L^t$ & $y_1$ & $y_2$ & $\pm\half + y_1 + y_2$ \\
$U_R^t$ & $y_1+\half$ & $y_2$ & $\half + y_1 + y_2$ \\
$D_R^t$ & $y_1+\half$ & $y_2-1$ & $-\half + y_1 + y_2$ \\
\hline
$T_L^b$ & $z_1$ & $z_2$ & $\pm\half + z_1 + z_2$ \\
$U_R^b$ & $z_1-\half$ & $z_2+1$ & $\half + z_1 + z_2$ \\
$D_R^b$ & $z_1-\half$ & $z_2$ & $-\half + z_1 + z_2$ \\
\hline
\end{tabular}
\end{center}
\end{table}

\section{Natural Topcolor-Assisted Technicolor}\label{sec:natural}

The questions of isospin violation and naturalness raised by CDT were
addressed by Eichten and me.~\cite{tctwoklee}$^{,\ts}$\cite{tctwokl} We
proposed that {\it different} technifermion isodoublets, $T^t$ and $T^b$,
give ETC mass to the top and bottom quarks. These doublets then could have
different $\uone$ charges which, however, were isospin-conserving for the
right as well as left-handed parts of each doublet.~\footnote{While this
eliminates the large value of $\rho - 1$ that concerned CDT, there remain
small, $\CO(\alpha)$, contributions from $Z$-$Z'$ mixing.} Thus,
there was no reason why the $\uone$ coupling~$g'_1$ could not be large,
hence no need to fine-tune the $\suone$ coupling~$g_1$.

The fermion content of the ``natural TC2'' models is given in Table~1.  In
these models, quarks get their ETC mass by coupling to doublet
technifermions $T^{l,t,b}$ via the $SU(2)\otimes\uy$-invariant interactions
terms
\be\label{eq:qTTqone}
\ba{ll}
\CH_{\ol u_i u_j} = &{g^2_{ETC} \over {M^2_{ETC}}} \ts \ol q^l_{iL}
\gamma^\mu T^l_L \ts \ol U^l_R \gamma_\mu u_{jR}  \ts + \ts\ts {\rm h.c.}
\\ \\
\CH_{\ol d_i d_j} = &{g^2_{ETC} \over {M^2_{ETC}}} \ts \ol q^l_{iL}
\gamma^\mu T^l_L \ts \ol D^l_R \gamma_\mu d_{jR}  \ts + \ts\ts {\rm h.c.}
\\ \\
\CH_{\ol t t} = &{g^2_{ETC} \over {M^2_{ETC}}} \ts \ol q^h_L \gamma^\mu T^t_L
\ts \ol U^t_R \gamma_\mu t_R   \ts + \ts\ts {\rm h.c.} \\ \\
\CH_{\ol b b} = &{g^2_{ETC} \over {M^2_{ETC}}} \ts \ol q^h_L \gamma^\mu T^b_L
\ts \ol D^b_R \gamma_\mu b_R   \ts + \ts\ts {\rm h.c.}
\ea
\ee
Here, $g_{ETC}$ and $M_{ETC}$ stand for generic ETC couplings and gauge
boson mass matrices. Also, $q^l_{iL} = (u_i, d_i)_L$ for $i=1,2$ and $q^h_L
= (t,b)_L$. The technifermions are color-singlets and transform under
$SU(N)$ technicolor as the fundamental $(\Ntc)$. The assignments of the
$\uone$ and $\utwo$ hypercharges, $Y_1$ and $Y_2$, for the quarks and
technifermions are listed in Table~1 in terms of six parameters ($x_{1,2}$;
$y_{1,2}$; $z_{1,2}$). The strong $\uone$ couplings of the right and
left-handed technifermions are isospin symmetric.

The technifermion hypercharges $x_i,y_i,z_i$ were fixed by requiring that
all $U(1)$ gauge anomalies cancel and that the $U(1)$ gauge symmetries
permit ETC four-fermion terms to (i) produce quark hard masses
(Eq.~\ref{eq:qTTqone}), (ii) induce generational mixing, and (iii) give mass
to all Nambu-Goldstone bosons except those involved in the electroweak
Higgs process. Note that we did {\it not} exhibit an explicit ETC model to
generate the four-fermion terms. The best we can do at this stage is to
assume that all four-fermion operators allowed by the gauge symmetries
exist.

For the purposes of this paper, the most important part of determining the
hypercharges was the selection of a four-technifermion (4T) operator to
induce generational mixing. Working in a ``standard'' chiral-perturbative
ground state in which technifermion condensates are diagonal, $\condij
\propto \delta_{ij}$, we found four possible 4T operators: 
\be\label{eq:TTTTmix}
\ba{ll}
\CH_{lttb} = &{g^2_{ETC} \over {M^2_{ETC}}} \ts 
\ol T^l_L\gamma^\mu T^t_L \ts \ol D^t_R \gamma_\mu D^b_R
\ts + \ts\ts {\rm h.c.} \\ \\
\CH_{bttl} = &{g^2_{ETC} \over {M^2_{ETC}}} \ts
\ol T^b_L\gamma^\mu T^t_L \ts \ol D^t_R \gamma_\mu D^l_R
\ts + \ts\ts {\rm h.c.} \\ \\
\CH_{lbbt} = &{g^2_{ETC} \over {M^2_{ETC}}} \ts 
\ol T^l_L\gamma^\mu T^b_L \ts \ol U^b_R \gamma_\mu U^t_R
\ts + \ts\ts {\rm h.c.}  \\ \\
\CH_{tbbl} = &{g^2_{ETC} \over {M^2_{ETC}}} \ts
\ol T^t_L\gamma^\mu T^b_L \ts \ol U^b_R \gamma_\mu U^l_R
\ts + \ts\ts {\rm h.c.}
\ea
\ee
These operators have the potential to induce the transitions $b_R \llra
s_L, d_L$; $\ts b_L \llra s_R, d_R$; $\ts t_R \llra c_L, u_L$; and $t_L
\llra u_R, c_R$, respectively.

The mixing that we do know about between the third and the first two
generations is contained in the KM matrix for left-handed quarks. It is
$|V_{cb}| \simeq |V_{ts}| \simeq$~0.03--0.05 $\sim m_s/m_b$ and $|V_{ub}|
\simeq |V_{td}| \simeq $~0.002--0.015 $\sim \sin\theta_C \ts
m_s/m_b$.~\cite{pdg} A nonzero mixing term~$\delta m_{sb} \sim m_s$ in the
$\ol s_L b_R$ element of the quark mass matrix is needed to produce mixing
of this magnitude. Recalling the way in which quark masses are generated in
Eq.~\ref{eq:qTTqone}, only $\CH_{lttb}$ has the correct flavor and chiral
structure to generate this kind of $\delta m_{sb}$. In effect, this
operator induces a mixing in the technifermion condensates so that $\langle
\ol D^l_L D^b_R \rangle \ne 0$.

Requiring the operator $\CH_{lttb}$ and sufficient additional ones to give
needed masses to Nambu-Goldstone bosons, we obtained just two solutions for
the hypercharges:
\be\label{eq:asolns}
\ba{llll}
{\rm A:}\qquad & x_1 = -\half\ts, \quad & y_1 = 0\ts, \quad & z_1 =
\half\ts,\\
& x_2 = \half\ts, \quad & y_2 = 0\ts, \quad & z_2 = -\half \ts; \\ \\
{\rm B:}\qquad & x_1 = 0\ts, \quad & y_1 = -1\ts, \quad & z_1 = 1\ts,\\
& x_2 = 0\ts, \quad & y_2 = 1\ts, \quad & z_2 = -1 \ts.
\ea
\ee
Both these solutions are phenomenologically acceptable in the sense that
they permit nontrivial patterns of technifermion condensation, $\condij$;
all technipion and top-pion masses are nonzero; the charged top-pion is
heavier than the top quark, so that $t \ra \toppip b$ does not occur; and
there is no large $\pi_T - \pi_t$ mixing so that $t \ra \pi_T^+ b$ is not a
problem, even if charged technipions are lighter than the top
quark.~\cite{balaji}

Although the problem of $B_d-\ol B_d$ mixing raised by Kominis was not
known to us when we wrote Ref.~[11], the $U(1)$ symmetries of the model
discussed there allow only the operator $\CH_{lttb}$ inducing $b_R \llra
d_L,s_L$. Thus, $|D^d_{Lbd}| \gg |D^d_{Rbd}|$ and the $B_d-\ol B_d$ constraint
is satisfied automatically.

In our model,~\cite{tctwoklee} the mechanism of topcolor breaking was
left unspecified and all technifermions were taken to be
$\suone\otimes\sutwo$ singlets. Thus, the $b_R \llra d_L,s_L$ transition
had to be generated by an externally induced term $\delta M_{ETC}$ in the
ETC mass matrix which transforms as $(\ol 3, 3)$ under the color groups. We
then estimated
\be\label{delm}
|V_{cb}| \simeq |D^d_{Lsb}| \simeq {\delta m_{sb} \over {m_b}}
\simle {\delta m_{sb} \over {m_b^{ETC}}} \simeq {\delta M^2_{ETC} \over
{M^2_s}} \ts\ts,
\ee
where $m_b$ is the mass of the $b$-quark and $M_s$ is the mass of the ETC
boson that generates the strange-quark mass, $m_s$. In a walking
technicolor theory,~\cite{wtc} $M_s \simge 100\,\tev$. However, we expect
$\delta M_{ETC} = \CO(1\,\tev)$ because that is the scale at which topcolor
breaking naturally occurs. This gives $s$--$b$ mixing that is about {\it
300~times} too small.
This problem is addressed in the rest of this paper. I shall connect
generational mixing with topcolor and electroweak symmetry breaking, all of
them occurring through technifermion condensation.

\section{TC2 Breaking and Generational Mixing}\label{sec:heart}

The discussion in this section closely follows that in Ref.~[12] with
some updating of the model and its discussion.

\subsection{Gauge Groups}\label{subsec:gg}

The gauge groups of interest to us are
\be\label{eq:groups} \sutc \otimes \suone \otimes \sutwo \otimes \uone
\otimes \utwo \otimes SU(2) \ts\ts, \ee
where, for definiteness, I have assumed that the technicolor gauge group is
$\sutc$. To help prevent light ``axions'', all of these groups (except for
the electroweak $SU(2)$ and, possibly, parts of the $U(1)$'s) must be
embedded in an extended technicolor group, $\Getc$. I will not specify
$\Getc$. This difficult problem is reserved for the future. However, as
above, I will assume the existence of ETC-induced four-fermion operators
which are needed to break quark, lepton and technifermion chiral
symmetries. Of course, these operators must be invariant under the groups
in Eq.~\ref{eq:groups}.

\subsection{$\uone \otimes \utwo$ Breaking}\label{subsec:ubreaking}

In order that top-quark condensation occur without unnatural fine-tuning,
the extra $Z'$ resulting from $\uone\otimes\utwo$ breaking must have a mass
of at most a few~TeV. Unlike the simple model of Ref.~[11], the
topcolor-breaking scenario described here seems to require that $Z'$
couples strongly to light, as well as heavy, quarks and leptons. Then, two
conditions are necessary to prevent conflict with neutral current
experiments. First, there must be a $Z^0$~boson with standard electroweak
couplings to all quarks and leptons. To arrange this, there will be a
hierarchy of symmetry breaking scales, with $\uone\otimes\utwo \ra \uy$ at
1--2~TeV, followed by $SU(2) \otimes \uy \ra U(1)_{EM}$ at the lower
electroweak scale $\Lambda_{EW}$. Assuming that technicolor interactions
induce both symmetry breakdowns, the technifermions responsible for
$\uone\otimes\utwo \ra \uy$---call them $\psi_L$ and $\psi_R$---must belong
to a {\it vectorial} representation of $SU(2)$ and to a {\it
higher-dimensional} representation of $\sutc$.~\footnote{To simplify the
analysis, I make the minimal assumption that the $\psi_{L,R}$ are
electrically neutral $SU(2)$ singlets.} Technifermions responsible for
$SU(2)\otimes \uy$ breaking will be assumed to belong to fundamental
representations of $\sutc$.~\footnote{This is reminiscent of multiscale
technicolor~\cite{multi} but, there, both the higher and fundamental
representations participate in electroweak symmetry breaking. In the
present model, I shall assume that $\psi_{L,R}$ belong to the $\half \Ntc
(\Ntc - 1)$-dimensional antisymmetric tensor representation. I also assume
that this set of technifermions is large enough to ensure that the
technicolor coupling ``walks'' for a large range of momenta.~\cite{wtc}}

The second constraint is that the $Z'$ should not induce large
flavor-changing interactions. This may be achieved if the $\uone$
couplings of the two light generations are GIM-symmetric. Then
flavor-changing effects will nominally be of order $|V_{ub}|^2/M^2_{Z'}$
for $\Delta B_d = 2$ processes, $|V_{cb}|^2/M^2_{Z'}$ for $\Delta B_s = 2$,
and negligibly small for $\Delta S = 2$. These should be within
experimental limits.

\subsection{$SU(3)_1\otimes SU(3)_2$ and Electroweak Breaking and
Generational Mixing}\label{subsec:main}

The fact that $b_R$ transforms as $(3, 1, 1; -\third)$ under $\suone
\otimes \sutwo \otimes SU(2) \otimes \uy$ while $d_L, s_L$ transforms as
$(1, 3, 2; \sixth)$ suggests that the mechanism connecting $d_L,s_L$ to
$b_R$ may also be responsible for breaking $\suone \otimes \sutwo \ra \suc$
{\it and} $SU(2)\otimes \uy \ra U(1)_{EM}$. Since the generational mixing
term transforms as $(\ol 3, 3)$ under the color groups, I introduce colored
technifermion isodoublets transforming under $\sutc \otimes \suone \otimes
\sutwo \otimes SU(2)$ as follows:
\be\label{eq:tquarks}
\ba{lll}
T^1_{L(R)} = &\left(\ba{ll} U^1 \\ D^1 \ea \right)_{L(R)}
&\in (\Ntc, 3, 1, 2(1)) \\
T^2_{L(R)} = &\left(\ba{ll} U^2 \\ D^2 \ea \right)_{L(R)}
&\in (\Ntc, 1, 3, 2(1)) \ts . \\
\ea 
\ee
The transition $d_L, s_L \llra D^2_L \llra D^1_R \llra b_R$
occurs if the appropriate ETC operator exists {\it and} if the condensate
$\condab$ forms.

The patterns of technifermion condensation, $\condij$, that do occur depend
on the strength of the interactions driving them and on explicit chiral
symmetry breaking (4T)~interactions that determine the correct
chiral--perturbative ground state, i.e., ``align the
vacuum''.~\cite{vacalign} The strong interactions driving technifermion
condensation are $\sutc$, $\suone$ and $\uone$. Technicolor does not prefer
any particular form for $\condij$; $\suone$ drives $\condaa \neq 0$;
$\uone$ drives $\condaa, \ts\ts \condbb \neq 0$ {\it or} $\condab \neq 0$,
depending on the strong hypercharge assignments.


\begin{table}[t]
\caption{Lepton, quark and technifermion colors and
hypercharges in the model of Sec.~5.
\label{tab:tabl2}}
\vspace{0.4cm}
\begin{center}
\begin{tabular}{|c|c|c|c|c|}
\hline
Particle & 
$SU(3)_1$ & 
$SU(3)_2$ & 
$Y_1$ & 
$Y_2$ \\
\hline
\hline
$\ell_L$ & $1$ & $1$ & $a$ & $-\half -a $ \\
$e_R$, $\mu_R$, $\tau_R$ & $1$ & $1$ & $a$ & $-1-a$ \\
\hline
$q_L^l$ & $1$ & $3$ & $b$ & $\sixth - b$ \\
$c_R$, $u_R$ & $1$ & $3$ & $b'$ & $\twothirds-b'$ \\
$d_R$, $s_R$ & $1$ & $3$ & $b''$ & $-\third-b''$ \\
\hline
$q_L^h$ & $3$ & $1$ & $d$ & $\sixth - d$ \\
$t_R$ & $3$ & $1$ & $d'$ & $\twothirds-d'$ \\
$b_R$ & $3$ & $1$ & $d''$ & $-\third-d''$ \\
\hline
$T_L^1$ & $3$ & $1$ & $u_1$ & $u_2$ \\
$U_R^1$ & $3$ & $1$ & $v_1$ & $v_2+\half$ \\
$D_R^1$ & $3$ & $1$ & $v_1$ & $v_2-\half$ \\
\hline
$T_L^2$ & $1$ & $3$ & $v_1$ & $v_2$ \\
$U_R^2$ & $1$ & $3$ & $u_1$ & $u_2+\half$ \\
$D_R^2$ & $1$ & $3$ & $u_1$ & $u_2-\half$ \\
\hline
$T_L^q$ & $1$ & $1$ & $w_1$ & $w_2$ \\
$U_R^q$ & $1$ & $1$ & $w'_1$ & $w'_2+\half$ \\
$D_R^q$ & $1$ & $1$ & $w'_1$ & $w'_2-\half$ \\
\hline
$T_L^l$ & $1$ & $1$ & $x_1$ & $x_2$ \\
$U_R^l$ & $1$ & $1$ & $x_1$ & $x_2+\half$ \\
$D_R^l$ & $1$ & $1$ & $x_1$ & $x_2-\half$ \\
\hline
$T_L^t$ & $1$ & $1$ & $y_1$ & $y_2$ \\
$U_R^t$ & $1$ & $1$ & $y'_1$ & $y'_2+\half$ \\
$D_R^t$ & $1$ & $1$ & $y'_1$ & $y'_2-\half$ \\
\hline
$T_L^b$ & $1$ & $1$ & $z_1$ & $z_2$ \\
$U_R^b$ & $1$ & $1$ & $z'_1$ & $z'_2+\half$ \\
$D_R^b$ & $1$ & $1$ & $z'_1$ & $z'_2-\half$ \\
\hline
$\psi_L$ & $1$ & $1$ & $\xi$ & $-\xi$ \\
$\psi_R$ & $1$ & $1$ & $\xi'$ & $-\xi'$ \\
\hline
\end{tabular}
\end{center}
\end{table}


In the approximation that technicolor interactions dominate condensate
formation, so that
\be\label{eq:cij}
\condij = -\half \Delta_T U_{ij} \qquad (i,j = 1,2) \ts,
\ee
the following is easily proved: If $T^1 \in (3, 1)$ and $T^2 \in (1, 3)$
are the only technifermions and if the vacuum-aligning interactions are
$\suone \otimes \sutwo$ symmetric then, in each charge sector, the unitary
matrix $U_{ij} = \delta_{ij}$ or $U_{ij} = (i\sigma_2)_{ij}$, but {\it not}
a nontrivial combination of the two. The diagonal form of $U$ is needed to
ensure that certain technipions get mass. The nondiagonal form is needed to
break topcolor and mix the heavy and light generations (with $\delta m_{sb}
\sim \langle \ol T^1 T^2 \rangle_{M_s}/M^2_s \sim m_s$). Therefore, in
order that both types of symmetry breaking occur, i.e., $U = a_0 1 + i a_2
\sigma_2$, it will be necessary to introduce additional technifermions to
complicate the vacuum alignment. We shall take these to be $(N,1,1,2(1))$
under $\sutc \otimes \suone \otimes \sutwo \otimes SU(2)$.

\subsection{A New Model}\label{subsec:model}

The model presented here is an improvement on the one published in
Ref.~[12] and presented at SCGT~96. That model had a strong axial-vector
coupling of the electron to the $Z'$ boson. This is excluded by
measurements of parity violation in cesium atoms unless $M_{Z'} \simge
20\,\tev$.~\cite{rscjt} But such a large $Z'$~mass cannot play an important
role in top-quark condensation without unnatural fine-tuning. Therefore,
the model of Ref.~[12] must be rejected.

In the new model, the electron has a purely vectorial coupling to the $Z'$
so that its mass can be as low as 2--3~TeV and still be consistent with
precision electroweak measurements.~\cite{jt} Unfortunately, this model and
all other models of this type that I have constructed have at least one
extra triplet of massless Nambu-Goldstone bosons in addition to the
longitudinal $W^\pm_L$, $Z^0_L$.~\footnote{The same was true of the
rejected model of Ref.~[12]. Since SCGT~96, I have realized that the
argument I gave there that there are no extra Nambu-Goldstone bosons is
wrong.}

The fermions in the new model, their color representations and $U(1)$
charges are listed at the end of this paper in Table~2. A number of choices
have been made at the outset to limit and simplify the charges and to
achieve the symmetry-breaking scenario's objectives:

\begin{itemize}

\item In order that electric charge is conserved by the technifermion
condensates, $u_1 + u_2 = v_1 + v_2$, $w_1 + w_2 = w'_1 + w'_2$, $y_1 + y_2
= y'_1 + y'_2$, and $z_1 + z_2 = z'_1 + z'_2$.

\item The $\uone$ charges of technifermions respect custodial isospin.

\item The most important choice for our scenario is that of the $\uone$
charges of $T^1$ and $T^2$. So long as $u_1 \neq v_1$, the broken $\uone$
interactions favor condensation of $T^1$ with $T^2$. If this interaction is
stronger than the $\suone$--attraction for $T^1$ with itself and if we
neglect other vacuum-aligning ETC interactions, then $\condij \propto (i
\sigma_2)_{ij}$ in each charge sector. In the extreme walking-technicolor
limit that the anomalous dimensions of $\condij$ are equal to one, this
condition is $\alpha_{Z'} (u_1 - v_1)^2 > 4\alpha_{V_8}/3$, where
$\alpha_{Z'}=g^{\prime 2}_1/4\pi$ and $\alpha_{V_8}=g^2_1/4\pi$. The proof
of this is presented in Ref.~[12].

\item I must choose $Y_1(t_R) = d' \neq Y_1(b_R) = d''$ to prevent strong
$b$-condensation.  For simplicity, I took $Y_1 = b' = b''$ for all
right-handed light quarks.  We shall see that solutions to the hypercharge
equations allow $dd'$ to be positive and greater than $dd''$, as it must be
in order that top quarks condense and bottom quarks don't.

\item For the $\sutc$ antisymmetric tensor $\psi$, $\xi' \neq \xi$
guarantees $\uone \otimes \utwo \ra \uy$ when $\langle \ol \psi_L \psi_R
\rangle$ forms. Note that, if $\Ntc = 4$, a single real $\psi_L$ is
sufficient to break the $U(1)$'s. Otherwise, to limit the parameters, $\xi'
= -\xi$ may be assumed.

\end{itemize}

To give mass to quarks and leptons, I assume the ETC operators:
\be\label{eq:qTTq}
\ba{ll}
\ol \ell_{iL} \gamma^\mu T^l_L \ts \ol D^l_R \gamma_\mu e_{jR}
\qquad &\Lra \qquad a - a' = x_1 - x'_1 = 0 \\ \\
\ol q^l_{iL} \gamma^\mu T^q_L \ts \ol T^q_R \gamma_\mu q^l_{jR} 
\qquad &\Lra \qquad b - b' = b - b'' = w_1 - w'_1 \\ \\
\ol q^h_L \gamma^\mu T^t_L \ts \ol U^t_R \gamma_\mu t_R
\qquad &\Lra \qquad d - d' = y_1 - y'_1 \\ \\
\ol q^h_L \gamma^\mu T^b_L \ts \ol D^b_R \gamma_\mu b_R
\qquad &\Lra\qquad d - d'' = z_1 - z'_1 \ts\ts. \\
\ea
\ee
The first condition guarantees a vectorial electron coupling to the
$Z'$, up to small mixing effects. To generate $d_L,s_L \llra b_R$, I
require the operator
\be\label{eq:sLbR}
\ol q^l_{iL} \gamma^\mu T^2_L \ts \ol D^1_R \gamma_\mu b_R
\qquad \Lra \qquad b - d' = 0 \ts\ts.
\ee
To suppress $d_R,s_R \llra b_L$, ETC interactions must not generate the
operator  $\ol q^h_L \gamma^\mu T^1_L \ts \ol D^2_R \gamma_\mu d_{iR}$.
This operator is forbidden by $U(1)$ interactions if
\be\label{eq:dbp}
d-b' \neq 0 \ts.
\ee
This constraint will turn out to follow from the required existence of
other four-fermion operators and certain no-anomaly constraints. Thus, this
operator does not appear without the intervention of $\uone$ breaking and
so the transition $d_R,s_R \llra b_L$ is automatically suppressed
relative to $d_L,s_L \llra b_R$ by a factor of $\delta M^2_{ETC}/M^2_s =
\CO(10^{-4})$. There should be no problem in this model with $B_d-\ol B_d$
mixing.~\cite{kominis}

The longitudinal components of the weak bosons will be the model's only
massless Nambu-Goldstone bosons if ETC and $\uone$ interactions (including
the operator $\ol q^h_L \gamma^\mu T^t_L \ts \ol U^t_R \gamma_\mu t_R$)
explicitly break all spontaneously broken chiral symmetries except
$SU(2)\otimes\uy$. For this, it is necessary that every $T_L^i$ {\it and}
$T_R^i$ appear in at least one term of the chiral-symmetry-breaking
Hamiltonian ${\cal {H}'}$.~\footnote{The proof of this statement---{\it
which applies to the ${\cal {H}'}$ obtained} after {\it vacuum
alignment}---makes use of the fact that the vector flavor symmetry charges
$Q^a$ annihilate the standard chiral-perturbative vacuum $|\Omega\rangle$,
and, so, axial charges $Q_5^a$ may be replaced by left- and/or right-handed
charges in using Dashen's theorem.~\cite{vacalign} If $T_L^i$ appears only
in $SU(2)\otimes\uy$-invariant terms of the form $\ol T_L^i \gamma_\mu
T_L^i \cdots$, there will be three additional NGBs, two charged and one
neutral. If $T_R^i$ appears only in terms of the form $\cdots \ol T_R^i
\gamma_\mu (a + b \sigma_3) T_R^i$, with $b \ne 0$, the charged NGBs acquire
mass, but the neutral one remains massless.}

I have not yet succeeded in building a model which allows enough 4T
operators to break all chiral symmetries. In the model presented here, one
maximal set of operators allowed by $\uone$ and $\utwo$ interactions and
consistent with the condition $u_1 - v_1 \ne 0$ required for $\condab \ne
0$ is:
\be\label{eq:TTTT}
\ba{ll}
\ol T^1_L \gamma^\mu T^t_L \ts \ol T^t_R \gamma_\mu T^1_R \qquad
&\Lra \qquad y_1 - y'_1  = u_1 - v_1 \\ \\
\ol T^2_L \gamma^\mu T^l_L \ts \ol T^t_R \gamma_\mu T^2_R \qquad
&\Lra \qquad y'_1 - x_1  = u_1 - v_1 \\ \\
\ol T^l_L \gamma^\mu T^t_L \ts \ol T^b_R \gamma_\mu T^l_R \qquad
&\Lra \qquad y_1 - z'_1  = y_2 - z'_2 = 0 \\ \\
\ol T^q_L \gamma^\mu T^t_L \ts \ol T^l_R \gamma_\mu T^q_R \qquad
&\Lra \qquad w_1 - w'_1  = y_1 - x_1 \ts\ts. \\
\ea
\ee
Note that these operators imply the equal-charge conditions
\be\label{eq:eqq}
x_1 + x_2 = y_1 + y_2 = z_1 + z_2 \ts.
\ee
These operators (and any generated by diagonal ETC interactions or $Z'$
exchange) leave $\sum_{i\ne b} \ol T^i_L \gamma_\mu \sigma^a T^i_L$ and
$\ol T^b_L \gamma_\mu \sigma^a T^b_L$ separately conserved.

The requirement that gauge anomalies cancel constrains $U(1)$ charge
assignments. Taking account of the equal-charge conditions, there are
five independent conditions which are linear in the hypercharges:
%
\be\label{eq:linanom}
\ba{ll}
\underline{U(1)_{1,2} [\sutc]^2:} &\\ \\
\qquad w_1 - w'_1 + y_1 - y'_1 + z_1 - z'_1
= -\half (\Ntc -2) (\xi - \xi') &\\ \\
{\underline{U(1)_{1,2} [\suone]^2:}} &\\ \\
\qquad 2d - d' - d'' = -2\Ntc(u_1-v_1) &\\
{\underline{U(1)_{1,2} [\sutwo]^2:}} &\\ \\
\qquad 2b - b'-b'' = \Ntc(u_1-v_1) &\\ \\
{\underline{U(1)_{1,2} [SU(2)]^2:}} &\\ \\
\qquad 3(a+2b+d) = -\Ntc [3(u_1 + v_1) + w_1 + x_1 + y_1 + z_1] &\\
\hskip1.109truein =\ts\ts\ts\ts\ts\Ntc [3(u_2 + v_2) + w_2 + x_2 + y_2 +
z_2] \ts\ts.&\\ \\
\ea
\ee
There are four anomaly conditions that are cubic in the
hypercharges. However, the $\uy[SU(2)]^2$ anomaly cancellation
guarantees that the $[\uy]^3$ anomaly also cancels, leaving three
independent conditions. Their most convenient form is:
\be\label{eq:cubnom}
\ba{ll}
{\underline{[U(1)_1]^3:}} &\\ \\
 \quad 0= 3[a^3 + 2(2 b^3 - b^{\prime 3} - b^{\prime\prime 3})
+ 2 d^3 - d^{\prime 3} - d^{\prime\prime 3}] &\\ 
\qquad + \half \Ntc(\Ntc - 1)(\xi^3 - \xi^{\prime 3}) 
 + 2\Ntc(w^3_1 - w^{\prime 3}_1 + y^3_1 - y^{\prime 3}_1 
 + z^3_1 - z^{\prime 3}_1 ) &\\ \\
{\underline{[\uone]^2 \uy:}} &\\ \\
\quad 0 = 2(b^2 - 2 b^{\prime 2} + b^{\prime\prime 2})
+ d^2 - 2 d^{\prime 2} + d^{\prime\prime 2} &\\
 \qquad +  2\Ntc [(w_1 + w_2) (w^2_1 - w^{\prime 2}_1)
+ (y_1 + y_2) (y^2_1 - y^{\prime 2}_1)
+ (z_1 + z_2) (z^2_1 - z^{\prime 2}_1)] &\\ \\
{\underline{[\uone]^3 +  [\utwo]^3 - 3[\uone]^2 \uy:}} &\\ \\
 \quad 0 = 2\Ntc\bigl\{(w'_1 - w_1) \ts[(w_1+w_2)^2 +\fourth] 
+ (y'_1 - y_1) \ts[(y_1+y_2)^2 +\fourth] &\\
 \qquad \ts\ts + (z'_1 - z_1) \ts[(z_1+z_2)^2 +\fourth]\bigr\}
+ 2(b' - b) + d'-d \ts\ts. &\\ 
\ea
\ee

Together with Eqs.~\ref{eq:qTTq}, \ref{eq:sLbR} and \ref{eq:TTTT}, the
linear anomaly conditions imply
\be\label{eq:relations}
\ba{ll}
& \Ntc = 4 \ts\ts \\ \\
& b-b' = b-b'' = w_1 - w'_1 = y_1-x_1 = \half\Ntc(u_1 - v_1) \\ \\
& d-d' = y_1 - y'_1 = u_1-v_1 \\ \\
& d-d'' = z_1 - z'_1 = -(2\Ntc + 1)(u_1 - v_1) \\ \\
& (\Ntc - 2) (\xi - \xi') = 3\Ntc (u_1 - v_1) \ts\ts. \\
\ea
\ee
The surprising requirement that $\Ntc = 4$ follows from $u_1-v_1 = y'_1-x_1
= y'_1-y_1 + y_1-x_1 = -(u_1-v_1) + \half\Ntc(u_1-v_1)$. Note that $b' =
b''$ and $b = d''$ imply that $\ol q^h_L \gamma^\mu T^1_L \ts
\ol D^2_R \gamma_\mu d_{iR}$ is not allowed by $U(1)$ interactions:
\be\label{eq:bppd}
b''-d = b'-b + d''-d = \half(3 \Ntc + 1) (u_1-v_1) \ne 0 \ts,
\ee
so long as $u_1 - v_1 \ne 0$.

As in Ref.~[12], I sought numerical solutions to the cubic anomaly equations
as follows: Requiring $u = \half(u_1 - v_1) \neq 0$, I set $\xi' = - \xi =
3\Ntc u/(\Ntc - 2)$. Then, choosing values for $w_1$, $y_1$, $y_1 + y_2$
and $d$, I solved for $u$, $a$ and $w_1 + w_2$. As a fairly random example,
the choice $\Ntc = 4$ and $w_1 = 1.00$, $y_1 = 1.00$, $y_1 + y_2 = 0$
(which implies $(w_1+w_2) = \pm{\textstyle{7\over{16}}}$), and $d = 1.00$
lead to the solutions
\be\label{eq:bsolns}
\ba{ll}
& u = -0.02691 \ts,\ts\ts \quad a = -1.6983 \ts,\ts\ts d' = 1.0538 \ts,\ts\ts
d'' = 0.5156 \\
& (w_1 + w_2 = -\sqrt{{7\over{16}}}) \\
\ea
\ee
\be\label{eq:csolns}
\ba{ll}
& u = -0.2100 \ts,\ts\ts a = -6.0867 \ts,\ts\ts  d' = 1.4200 \ts,\ts\ts
d'' = -2.7796  \\
& (w_1 + w_2 = \sqrt{{7\over{16}}}) \\ \\
\ea
\ee
Note that $dd' > dd''$ for both solutions. The lower limits on $M_{Z'}$ for
the first solution is $M_{Z'} = 2.7\,\tev$.~\cite{jt} It is three times
larger for the second one, requiring fine-tuning of the $\uone$
coupling. Even worse, the value $a \simeq -6$ is so large that lepton
condensates should form! A more thorough search will turn up many more
solutions, some acceptable, some not. What is needed in this class of
models is a solution to the problem of massless technipions and, more
generally, a careful study of the vacuum alignment problem.

\section{Conclusions}\label{sec:conclude}

Topcolor-assisted technicolor is now the most promising {\it natural}
approach to a dynamical explanation for electroweak symmetry breaking and
all quark and lepton masses and mixing parameters. The most economical
scheme for breaking electroweak and topcolor gauge symmetries {\it and}
connecting the heavy and light generations would seem to involve
technicolor and topcolor interactions alone. Here, I have outlined a
scenario to implement that. As we have moved deeper into this scenario,
several hurdles have appeared in our course. More work is needed to know if
these are show-stoppers. Clever ideas are needed to advance the whole TC2
approach. More than anything, experimental data are needed to show us the
the way to a more complete picture. Some of the phenomenological questions
which experiments can address are discussed by Estia Eichten in his talk at
this workshop.~\cite{eescgt}

\section*{Acknowledgments}

I extend warm thanks to the SCGT~96 organizers, especially Koichi
Yamawki and Masako Bando, for an excellent workshop, gracious hospitality,
and the opportunity to visit Japan, see a few of its many sights, meet some
very nice people, and eat some really wonderful food. The research reported
here was supported in part by the Department of Energy under
Grant~No.~DE--FG02--91ER40676.

\end{document}